# Competing magnetic phases in *Ln*SbTe (*Ln* = Ho and Tb)

Igor Plokhikh[1,*], Vladimir Pomjakushin[2], Dariusz Jakub Gawryluk[1], Oksana Zaharko[2], Ekaterina Pomjakushina[1]

[1]Laboratory for Multiscale Materials Experiments, Paul Scherrer Institut, PSI, Villigen, CH-5232, Switzerland

[2]Laboratory for Neutron Scattering and Imaging (LNS), Paul Scherrer Institut, PSI, Villigen, CH-5232, Switzerland

[*]Corresponding author: igor.plokhikh@psi.ch; igor.plohih@gmail.com

## *Abstract*

The interplay between topological electronic structure and magnetism may result in intricate physics. In this work, we describe a case of rather peculiar coexistence or competition of several magnetic phases below seemingly single antiferromagnetic transition in *Ln*SbTe (*Ln* = Ho and Tb) topological semimetals, the magnetic members of the ZrSiS/PbFCl structure type (space group *P*4/*nmm*). Neutron diffraction experiments reveal a complex multi-step order below $T_N$ = 3.8 K (*Ln* = Ho) and $T_N$ = 6.4 K (*Ln* = Tb). Magnetic phases can be described using four propagation vectors: $k_1$ = (½ 0 0) and $k_2$ = (½ 0 ¼) at the base temperature of 1.7 K, which transform into incommensurate vectors $k_1'$ = (½ - δ 0 0), $k_3$ = (½ - δ 0 ½) at elevated temperatures in both compounds. Together with the refined models of magnetic structures, we present the group theoretical analysis of magnetic symmetry of the proposed solutions. These results prompt further investigations of the relation between the electronic structure of those semimetals and the determined antiferromagnetic ordering existing therein.

## *Introduction*

Topological state of matter is currently in the focus of solid-state research due to promising physical properties originated from the non-trivial electronic structure and magnetism.[1-5] Topological semimetals host exotic quasiparticles – relativistic fermions, the solid-state analogs of Dirac and Weyl fermions from particle physics. These semimetals may exhibit useful technological properties like large magnetoresistance and high charge carrier mobility.[6-8] Whereas, topological particle-like objects, skyrmions, are proposed to be useful for the next generation of memory storage devices.[9]

Square-net PbFCl-type (*P4/nmm*) non-symmorphic topological semimetals of the ZrSiS group along with its magnetic analogs *Ln*SbTe (*Ln* – lanthanide) attract attention because their band structure hosts Dirac nodes.[10] For the prototype, ZrSiS, the energy range of the linearly dispersed bands is over 2 eV above and is located below the Fermi level.[11] The magnetic *Ln*SbTe members, featuring topological electronic structure, charge-ordering instabilities and strong localized 4*f* magnetism, may serve as an ideal platform to study the interplay between electronic structure, nonsymmorphic crystal symmetry as well as magnetic correlations and frustration. Substantial structural flexibility allows to modify the electronic structure, symmetry (breaking the time reversal for the magnetic members) and charge ordering, leading to rich phase diagrams. For example, by partial substitution of Sb by Te in GdSbTe, it is possible to obtain $GdSb_{1-x}Te_{1+x}$ series of solid solutions, where depending on the value of *x*, different charge ordering patterns associated with different shapes of Fermi surface are observed.[12-14] One of the members, namely $GdSb_{0.46}Te_{1.48}$ was found to be a nearly ideal nonsymmorphic Dirac semimetal with a Dirac node at the Fermi level. It was proposed to host the antiferromagnetic skyrmion phase, which has not yet been confirmed.[15] Similar substitution can be done for CeSbTe, where it leads to evolution of complex magnetization curves (Devil's staircase magnetization) originated from perturbing of magnetic interactions along the *c*-axis.[16-18]

Despite the growing interest to CeSbTe and GdSbTe members of the *Ln*SbTe series, there are only fragmentary reports on physical properties related with topology, ARPES measurements and crystal structures of the other lanthanide members. The recent reports on HoSbTe[19-21] hint to a narrow band gap driven by spin-orbit coupling, a bad-metal-like state observed in resistivity measurements and an antiferromagnetic transition due to ordering of the $Ho^{3+}$ moments. The magnetism in this material is strongly anisotropic with the easy-axis within the tetragonal *a-b* plane. Besides, there are reports on electronic and crystal structure (including complex superstructures), magnetic and transport properties of LaSbTe[22,23], SmSbTe[24,25] and NdSbTe[26,27].

The aim of this work is to understand the nature of the magnetically ordered phases in magnetic *Ln*SbTe materials and to link it with the electronic structure. Surprisingly, below seemingly single Néel temperature in *stoichiometric* (1:1:1)

HoSbTe and TbSbTe, the thermal evolution of neutron diffraction pattern features a plethora of strong magnetic reflections, which cannot be described by the magnetic phase with single propagation vector at all temperatures and the reflections even have different temperature dependences within the antiferromagnetic state.

*Experimental part*

Homogeneous powder samples of $Ln$SbTe ($Ln$ = Ho or Tb) have been prepared from the elements (Ho 3N, Chempur, Tb 3N abcr, Sb 6N Alfa Aesar, Te 5N Alfa Aesar). The starting elements were mixed in 1:1:1 ratio in a He-glove box, pressed into pellets, sealed in silica ampules, slowly heated and annealed consequently at 800 ˚C (72 h heating, 24 h holding) and 900 ˚C (12 h heating, 24 h holding) with intermediate grinding. Sample purity was checked and the data for structure refinement were collected at room temperature using BRUKER AXS D8 ADVANCE diffractometer (CuKα radiation, Bragg-Brentano geometry, 1D LynxEye PSD detector, 5 – 120˚ 2θ range). DC magnetization of the samples was measured using MPMS XL 7 T, Quantum Design magnetometer in 1.8 K – 300 K temperature range and 0 – 7 T field range. Sintered polycrystalline pieced loaded in gelatin capsules were used for magnetization measurements.

Neutron powder diffraction data has been collected on the HRPT diffractometer[28] (SINQ, PSI) using the wavelength λ = 2.449 Å (Ge(400) monochromator) in 2θ range of 3.55–164.50˚, and steps of 0.05˚. The 5.6 g of HoSbTe and 7.5 g of TbSbTe were loaded in vanadium cans (10 mm inner diameter), mounted in a sample changer into an Orange He-cryostat (1.6 – 310K). For HoSbTe, diffraction patterns were collected at the following temperatures: 1.7 K, between 1.9 K and 4 K with 0.1 K temperature steps, at 5 K and at 10 K. For TbSbTe, diffraction patters were collected at 1.7 K, between 2 K and 6.5 K with 0.5 K steps (or finer around the transition-like peculiarities) and at 10 K.

X-ray and neutron diffraction datasets were analyzed using JANA2006/JANA2020 software using standard mathematical approach for description of powder diffraction profile (Pseudo-Voight profile function, Legendre polynomials for background).[29,30] Search of *k*-vectors was done using the *k*-search code, implemented in the FullProf suite.[31] Representation and magnetic symmetry analysis[32] have been done using the ISODISTORT tool from the ISOTROPY software[33,34] and some tools from the Bilbao crystallographic server (k-SUBGROUPSMAG, MAXMAGN).[35,36] Magnetic structures were plotted using the VESTA visualization tool.[37]

The primary structure characterization was performed with the laboratory X-ray diffractometer. The crystal structure of HoSbTe is consistent with the one reported in the literature. The structure of a new series member TbSbTe was also well refined using the same model. The summary of the crystal structure parameters refined from X-ray powder diffraction data is given in **Table 1**. No impurity phases were detected. In addition, although Te and Sb are poorly distinguishable using X-ray diffraction, the crystal structure refined against neutron powder diffraction data in paramagnetic state (*vide infra*) indicate, that both compounds are stoichiometric (1:1:1).

*Table 1.* Summary of crystal structure refinements for *Ln*SbTe (*Ln* = Tb and Ho). Space group *P*4/*nmm* (No. 129, origin choice 1). The coordinates and Wyckoff positions of atoms are: *Ln* (2*c*, ½ 0 *z*), Te(2*c*, 0 ½ *z*), Sb(2*a*, 0 0 0).

| Compound | HoSbTe | TbSbTe |
|---|---|---|
| $a, c$, Å | 4.2299(1), 9.1524(1) | 4.2506(1), 9.2178(1) |
| $z(Ln)$, $z(Te)$ | 0.2762(1), 0.3761(1) | 0.27585(6), 0.37562(7) |
| $d(Ln – Te)$, $d(Ln – Sb)$ $d(Sb – Sb)$, Å | 3.1277(4), 3.2960(9) 2.9910(1) | 3.1432(2), 3.3140(5) 3.0057(1) |
| $U_{iso}(Ln)$, $U_{iso}(Sb)$, $U_{iso}(Te)$ in $10^4 \cdot Å^2$ | 59(5), 9(4), 30(4) | 74(3), 29(3), 54(3) |
| $R_P$, $R_{WP}$, $R_I$ in %, $\chi^2$ | 3.6, 4.7, 1.5, 1.2 | 1.5, 2.1, 1.8, 1.8 |

### Results and discussion

*Magnetization data*

*Ln*SbTe (*Ln* = Ho and Te) are the magnetic members of the ZrSiS (CeSbTe) family. The magnetic properties for both samples are summarized in **Figures 1** and **2**. Above 50 K both samples behave like Curie-Weiss paramagnets with $\theta_{CW} = -7.1(2)$ K and $\mu_{eff} = 10.67(2)$ $\mu_B$ for *Ln* = Ho and $\theta_{CW} = -17.1(2)$ K and $\mu_{eff} = 9.76(2)$ $\mu_B$ for *Ln* = Tb. While the values of magnetic moments are consistent with free $Ln^{3+}$ values (10.6 $\mu_B$ for $Ho^{3+}$ and 9.7 $\mu_B$ for $Tb^{3+}$), the negative Curie-Weiss temperatures indicate predominantly antiferromagnetic interaction between the $Ln^{3+}$ magnetic centers in the paramagnetic temperature region. Indeed, the low-field (0.01 T) magnetic susceptibility curves measured in the zero-field cooling mode feature antiferromagnetic transitions at $T_N$ = 3.8 K for HoSbTe and $T_N$ = 6.4 K for TbSbTe. Note that field-cooling and zero-field cooling curves measured in 0.01 T nearly coincide. The value of $T_N$ for HoSbTe is consistent with the published data, where the single antiferromagnetic transition has been deduced through magnetization, heat capacity and resistivity measurements. The Néel temperatures in both compounds are lower than the absolute values of the Curie-Weiss temperatures indicating a certain degree of magnetic frustration, probably driven by the anisotropic 2D character of the title structure.

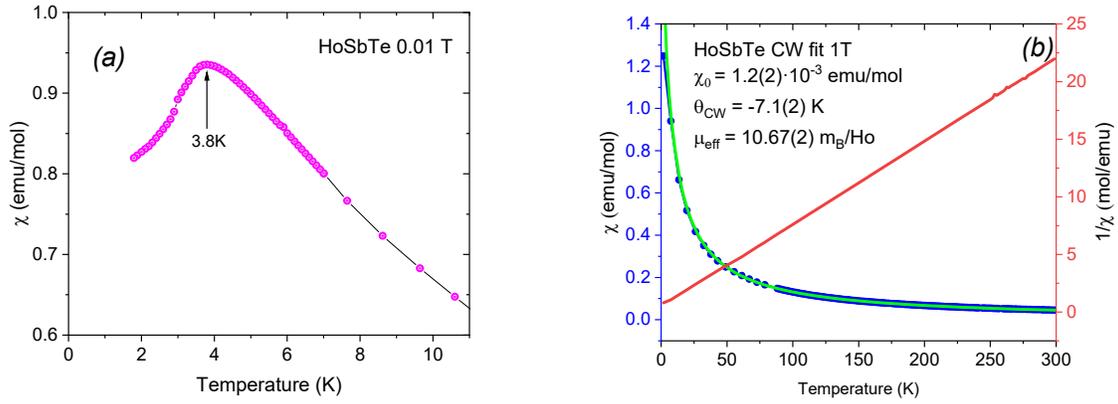

*Figure 1.* (*a*) magnetic susceptibility for HoSbTe in 0.01 T at low temperature outlining the magnetic phase transition at $T_N = 3.8$ K and (*b*) magnetic susceptibility in 1 T, inverse magnetic susceptibility and the Curie-Weiss fit.

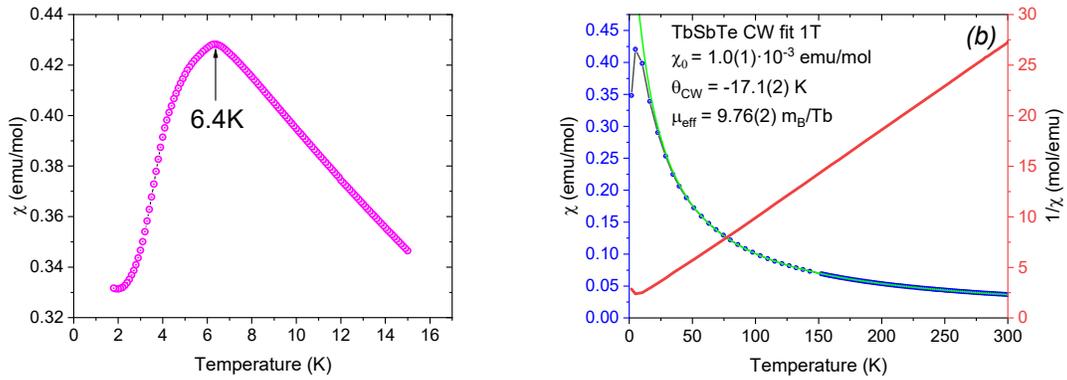

*Figure 2.* (*a*) magnetic susceptibility for TbSbTe in 0.01 T at low temperature outlining the magnetic phase transition at $T_N = 6.4$ K and (*b*) magnetic susceptibility in 1 T, inverse magnetic susceptibility and the Curie-Weiss fit.

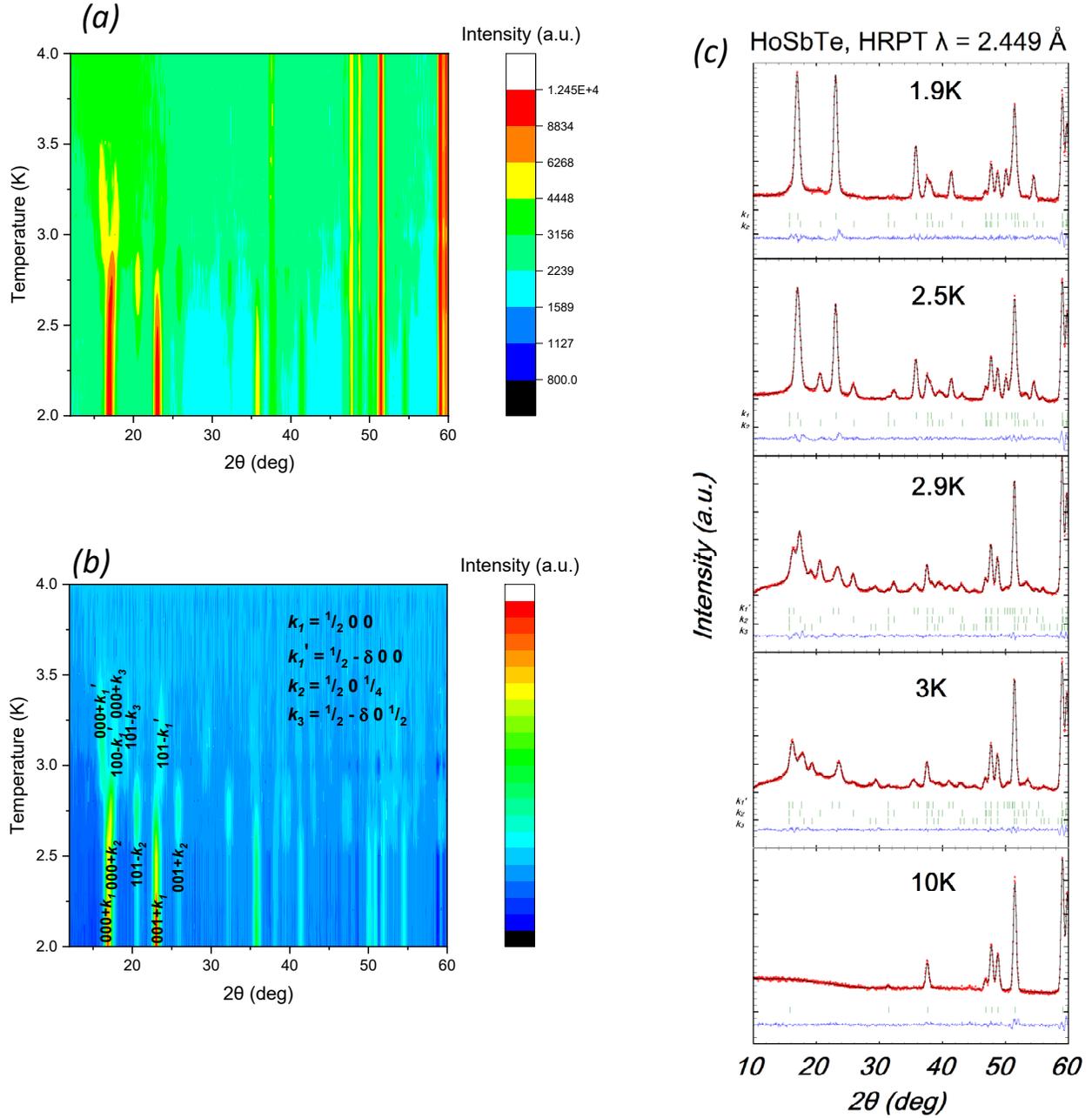

*Figure 3.* Neutron diffraction data for HoSbTe: (*a*) temperature evolution of overall scattering, (*b*) temperature evolution of magnetic scattering (overall scattering minus 5 K paramagnetic data), (*c*) Le Bail fit (as explained in the text) of diffraction patterns at selected temperatures (experimental points – red, calculated curve – black, difference curve – blue, positions of nuclear and magnetic reflections – green). The indexes of the strongest magnetic reflections of each propagation vector are provided on the magnetic scattering chart (*b*).

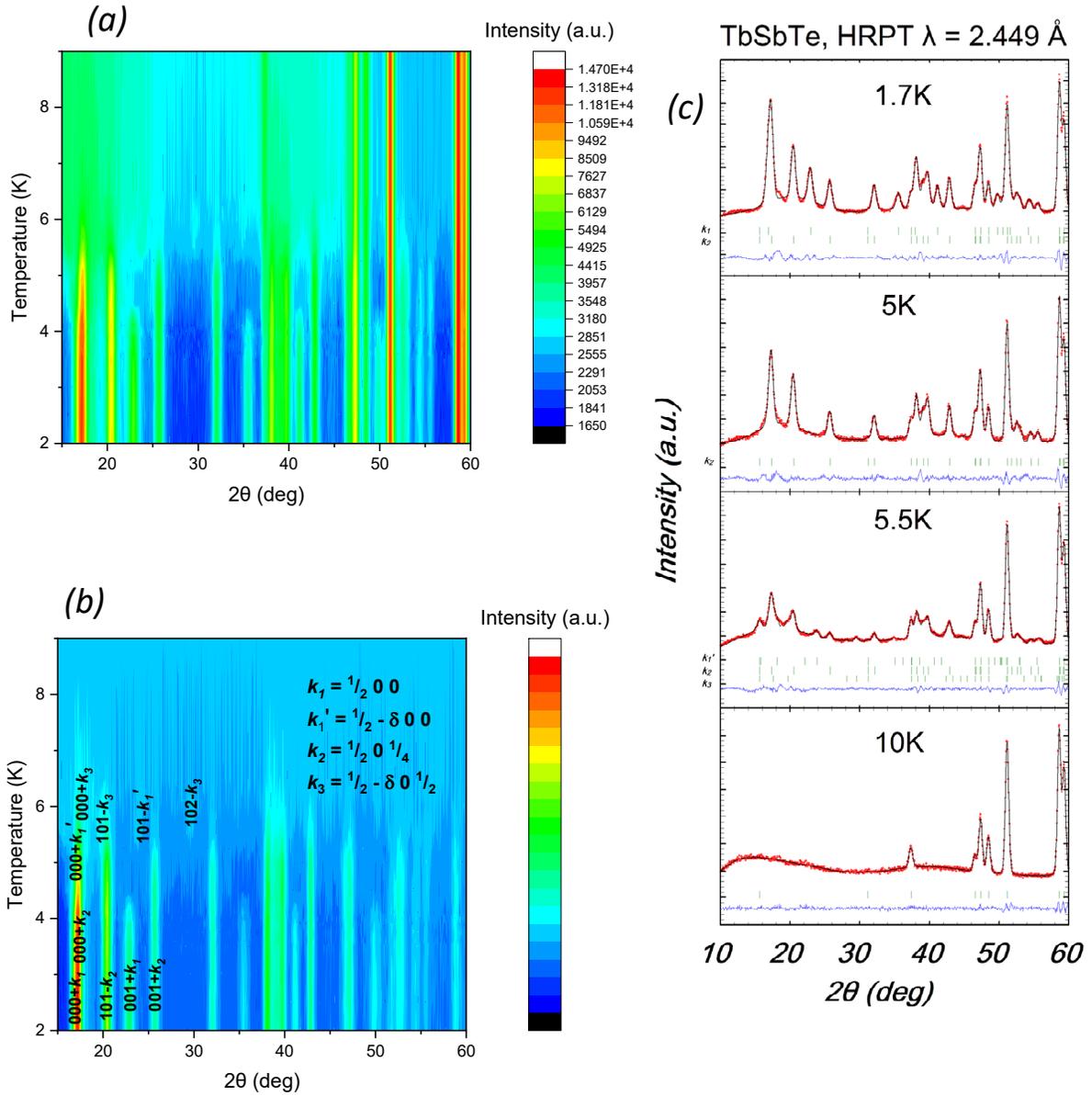

*Figure 4.* Neutron diffraction data for TbSbTe: (*a*) temperature evolution of overall scattering, (*b*) temperature evolution of magnetic scattering (overall scattering minus 10 K paramagnetic data), (*c*) Le Bail fit of diffraction patterns at selected temperatures (experimental points – red, calculated curve – black, difference curve – blue, positions of nuclear and magnetic reflections – green). The indexes of the strongest magnetic reflections of each propagation vector are provided on the magnetic scattering chart (*b*).

*Powder neutron diffraction. Indexing of magnetic reflections*

To characterize the magnetically ordered states microscopically, we have collected the neutron powder diffraction patterns in the relevant temperature ranges. The temperature dependences of the patterns are visualized in **Figures 3** and **Figure 4**, for *Ln* = Ho and *Ln* = Tb, respectively. The first step in the determination of a magnetic structure is the identification of the propagation vectors. This is done using so-called LeBail fitting, where all peak intensities are refined separately without any structure model, thus allowing to fit the propagation vectors and the crystal metrics. The 10 K diffraction pattern of HoSbTe can be fully described by nuclear reflections consistent with RT X-ray diffraction data; the 5 K diffraction pattern (not shown) additionally features a significant diffuse scattering indicative of approaching the magnetic phase transition. At the lowest temperature 1.7 K in HoSbTe, a set of magnetic reflections is observed, which can be indexed with the propagation vector $k_1 = (½\ 0\ 0)$. On increasing the temperature, the intensities of magnetic reflections corresponding to $k_1$ gradually decrease and they are nearly gone at 2.8 K. In this temperature range, a set of new magnetic reflections gains intensity; they all can be assigned to another propagation vector $k_2 = (½\ 0\ ¼)$. Being rather pronounced at 2.9 K, the intensities of $k_2$ reflections become very small at 3 K. Still, below 2.9 K both $k_1$ and $k_2$ are present, albeit their contents vary with temperature. The $k_1$ reflection at *ca.* $2\theta = 18.5\ °$ splits into another feature in between 2.8 K and 3 K; implying its transformation to its incommensurate counterpart, namely $k_1' = (½ - \delta\ 0)$. Indeed, using $\delta = 0.022(1)$ it is possible to describe the majority of magnetic intensities in the 3 K diffraction pattern in the whole angular range. However, there are few remaining reflections (for instance, at *ca.* $2\theta = 19.8\ °$), not present below 2.9 K and not corresponding to $k_1$, $k_1'$ or $k_2$. Using the *k*-search code implemented in the FullProf suit, we index them by a third propagation vector, $k_3 = (½ - \delta'\ 0\ ½)$. This vector also describes several weak reflections at yet higher $2\theta$ angles. Although from the first glance, there is no reason for $\delta$ and $\delta'$ to be equal, the constraining them to the same value does not harm the quality of fit.

As follows from **Figure 4**, the situation in TbSbTe is qualitatively similar. At the base temperature about 1.7 K, the main reflections of the propagation vectors $k_1 = (½\ 0\ 0)$ and $k_2 = (½\ 0\ ¼)$ are present. The $k_2$ reflections are stronger than in HoSbTe. Intensities of both *k*-vectors are nearly temperature independent up to 3.5 K, above which the intensities of $k_1$ start to decrease and are gone at 4.5 K.

Around 4 K the intensities of $k_2$ reflections start decreasing and are nearly gone at 5.5 K. Similar to HoSbTe, all magnetic intensities at 5.5 K can be described with three propagation vectors $k_2$, $k_1' = (½ – δ\ 0\ 0)$ and $k_3 = (½ – δ\ 0\ ½)$ with $δ = 0.035(1)$.

Schematic phase diagrams with symmetry of the ordered phases are provided in *Figure 5*. In the next two chapters, we suggest the magnetic structures for the obtained propagation vectors using both magnetic symmetry and irreducible representation arguments. We use here the internationally established nomenclature for the irreducible representations (irreps) labels, Shubnikov magnetic space groups (MSG) and magnetic superspace space groups (MSSG).[32–34, 35, 38] The detailed tables with numerical data for all refined magnetic structures together with plotted interactive models (VESTA files) are provided in supplementary information (SI).

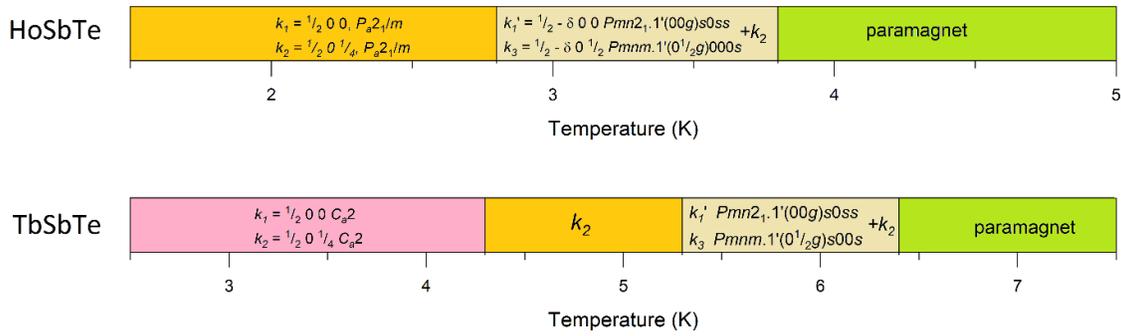

*Figure 5*. Schematic phase diagrams for *Ln*SbTe (*Ln* = Ho and Tb) within the magnetically ordered region. The space groups provided on the panels are explained in the next two chapters.

*Models of magnetic structures for HoSbTe*

The possible models of magnetic structures for HoSbTe for $k_1 = (½\ 0\ 0)$ were constructed using the representation analysis routine implemented in JANA2020 and with the ISODISTORT tool. The propagation vector $k_1$ has two arms ((½ 0 0) and (0 ½ 0)), corresponds to the X point of Brillouin zone (BZ) and yields two possible magnetic irreps: $mX_1$ and $mX_2$. Assuming that only one arm of the propagation vector is active each irrep generates two maximal symmetry subgroups as shown in *Figure S1*. The orthorhombic candidates ($P_amm2.1'$ and $P_ama2.1'$) can be immediately ruled out because they split the Ho site into two symmetrically independent ones and constrain the size of magnetic moment on one of them to zero. Contrary to that, the monoclinic MSGs $P_a2_1/m$ (BNS index

No. 11.55, basis = {(0,2,0),(-1,0,0),(0,0,1)})[1] and $P_c2_1/c$ (No. 14.82, basis = {(0,0,1),(1,0,0),(0,2,0)}), corresponding to $mX_1$ and $mX_2$, both feature a single Ho site in the unit cell. $mX_1$ with one variable (amplitude of magnetic moment along the *b*-axis) provides the best fit to the observed magnetic intensities.

Magnetic propagation vector $k_2$ = (½ 0 ¼) corresponds to the incommensurate W (0 ½ g) point, with g locked to the commensurate value ¼ in our case. W point has two complex irreducible representations $W_1$ and $W_2$, and two complex-conjugated to them irreps $W_3$ and $W_4$. The Herring coefficient[33,34,38] for both irreducible representations is H = 0 (or it is also defined as type 3), implying that $W_1$ and $W_3$ ($W_2$ and $W_4$) are not equivalent and have to be combined in pairs. Possible solutions were constructed using the ISODISTORT wizard. The W point in this case yields, as explained above, two physically irreducible magnetic representations $mW_1W_3$ and $mW_2W_4$. We note that in this special case, when we are forced to combine two non-equivalent representations, the kernel symmetry can be lower as compared to single irrep solutions. The above representations have the maximal symmetry one-arm solutions in the MSGs $P_a2_1/m$ (BNS index No. 11.55, basis={(0,2,0),(1,0,0),(0,-1,-2)}) and $P_c2_1/c$ (No. 14.82 basis = {(0,1,2),(-1,0,0),(0,-2,0)}), for $mW_1W_3$ and $mW_2W_4$, respectively. The first one provides a better fit to the diffraction intensities. This solution features two symmetrically independent Ho sites with variable amplitude of magnetic moment along the *b*-axis (relative to the parent cell), which can be constrained to be equal without harming the quality of fit.

The best candidates for $k_1$ and $k_2$ were simultaneously refined against the 2.5 K diffraction pattern to verify the validity of both models. The absence of significant deviations in the difference curve along with low profile residuals prove the correctness of both maximal symmetry models (see ***Figure 6*** and ***Figure 7***). Hereafter for conciseness, we refer to the magnetic structure models as $k_1$, $k_2$, $k_1'$, and $k_3$-structures, which correspond to the determined magnetic structures within the indicated in the text magnetic (super)space groups.

---

[1] The magnetic space group is defined by its symbol and the basis transformation, *i.e.* the matrix that transforms the parent nuclear cell to magnetic unit cell. Note, that although in some cases the space group symbols are the same, the basis can be different, leading to essentially different solution.

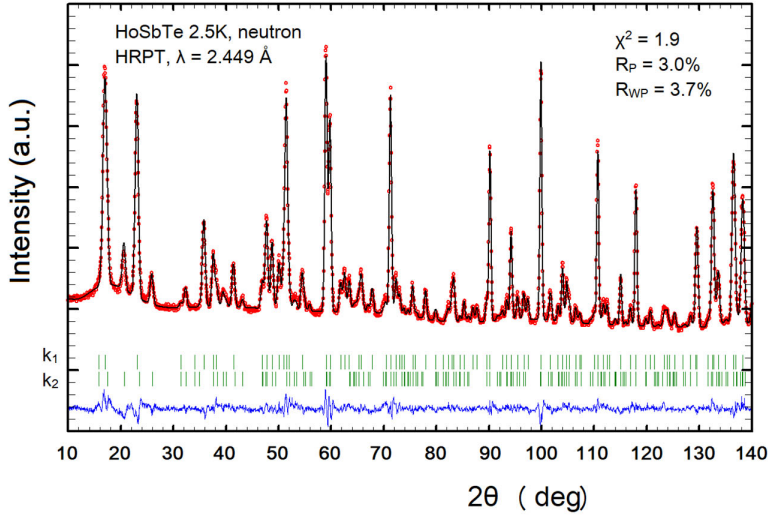

*Figure 6.* Experimental points (red), calculated curve (black), difference curve (blue) and positions of reflections (green) – Rietveld refinement plot for neutron data of HoSbTe at 2.5 K.

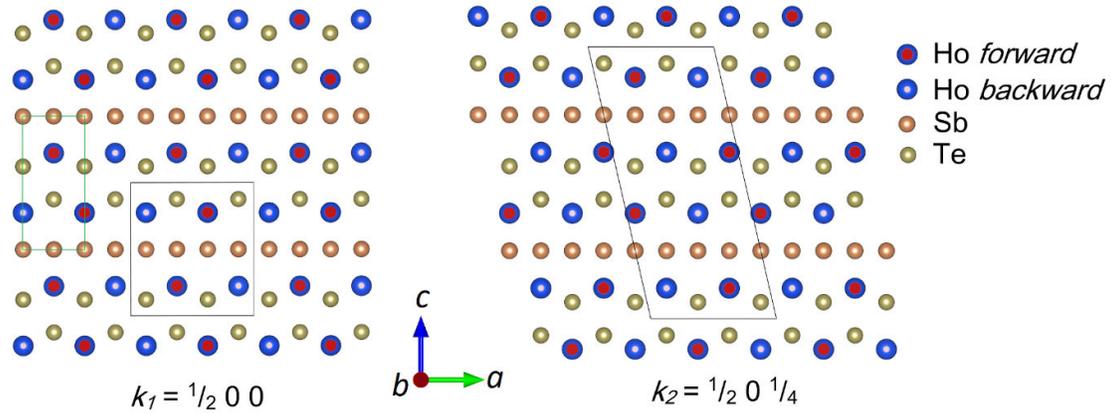

*Figure 7.* Projection of the refined $k_1$ and $k_2$ magnetic structures in HoSbTe at 2.5 K. The magnetic unit cells are outlined in black, the parent unit cell with Sb positioned at the origin is outlined in green. The coordinate axes are indicated for the parent cell.

The propagation vector $k_1' = (½ – δ\ 0\ 0)$ is the DT point with four possible irreducible representations, each of them has two possible solutions (the incommensurate 3D+1 magnetic subgroups of the *Pmmn* and *Pm2₁n* space groups, respectively). Testing these eight models against the diffraction intensities shows that only one MSSG *Pmn2₁*.1'(00g)s0ss (No. 31.1.9.2.m124.2) present for both mDT$_2$ and mDT$_4$ irreps, fits the experimental Bragg peaks. These two solutions, being non-centrosymmetric, represent mirror images of each other, and cannot be distinguished from our powder diffraction data. A similar situation is obtained for $k_3 = (½ – δ\ 0\ ½)$ vector, which is located at the U point of BZ. Here the MSSG $Pmnm.1'(0^1/_2g)000s$ (No. 59.1.10.6.m406.2) generated by the mU$_3$ irrep describes the experimental intensities well.

The coexistence of three propagation vectors ($k_1'$, $k_3$ and $k_2$), the first two of which are incommensurate, in diffraction pattern at 3 K makes the density of reflections per given 2θ region very high, even at low 2θ angles. This, together with significant diffuse scattering, limits the quantitative information that can be extracted from the available data, even though we used relatively long-wavelength neutrons. On the other hand, the computational issues preclude the possibility to refine via Rietveld method two incommensurate structures against a single diffraction pattern, as it has been done for commensurate $k_1$ and $k_2$. (**Figure 6**). Keeping this in mind and appreciating the limitations imposed, we performed the refinements for $k_1'$ and $k_3$ in two separate runs in the following way: a magnetic phase of the interest ($k_1'$ or $k_3$) was refined via Rietveld method, while two others were included using Le Bail algorithm. We consider this to be a valid approach because in the low angle region (below 2θ = 40°), there are strong reflections for both vectors at the positions that are not affected by the two remaining phases, and hence allow discriminating different magnetic models. The resulting profile matching plots are shown in **Figure 8** together with the derived magnetic models.

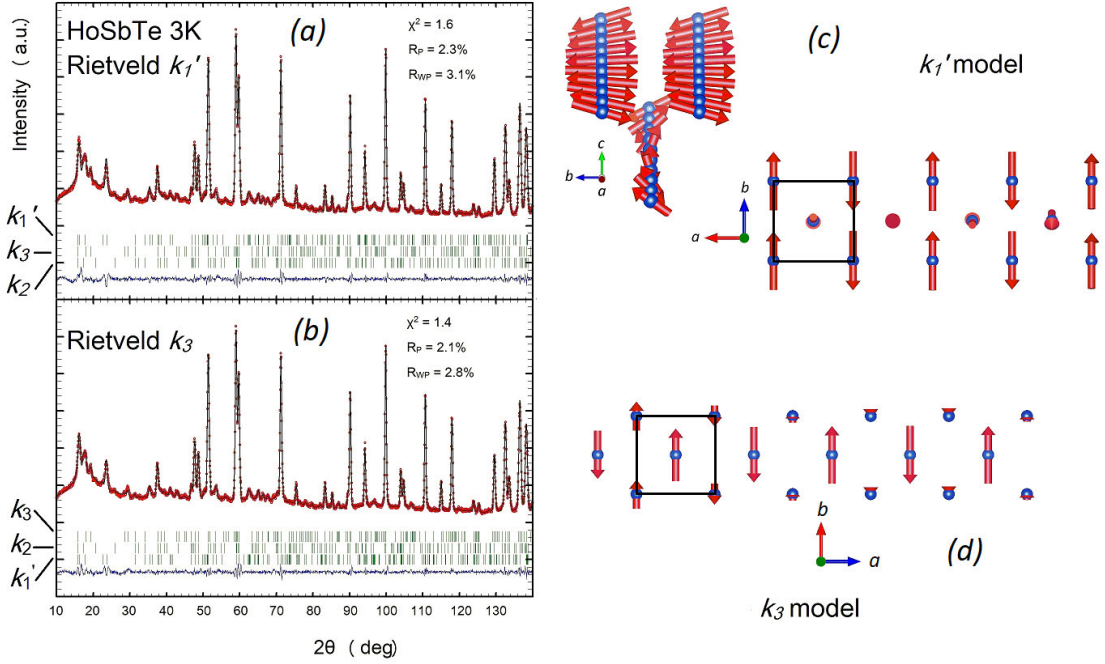

*Figure 8.* (left) neutron diffraction pattern at 3 K for HoSbTe refined using Rietveld method for the $k_1'$ model *(a)* and $k_3$ model *(b)*. The remaining vectors were included in refinement through Le Bail method (experimental points – red, calculated curve – black, difference curve – blue, positions of reflections – green). (right) the schematic representation of magnetic structures corresponding to $k_1'$ *(c)* and $k_3$ *(d)*.

The $k_1 = (½\ 0\ 0)$ and $k_2 = (½\ 0\ ¼)$ magnetic structures are collinear with magnetic moments along the *b*-axis (*Figure 7*). Sheets of Ho moments in the *a-b* plane are stack in the antiparallel manner along the *a*-axis and in the parallel manner along the *b*-axis. To form a [HoTe] slab (placed between two Sb-nets), two of such sheets are placed on top of each other with a (½, ½, z) shift. In the $k_1$ structure, such slabs are simply expanded along the *c*-axis through the lattice translations, forming on average an antiferromagnetic structure. The $k_2$ vector has a component along the *c*-axis; thus, two slabs along the *c*-axis are not transformed into each other through a lattice translation, but rather transform into each other through a glide plane (½a shift) passing through the Sb-net.

Both $k_1' = (½ - δ\ 0\ 0)$ and $k_3 = (½ - δ\ 0\ ½)$ magnetic structures are incommensurate and conceptually very similar to $k_1$ and $k_2$, with magnetic moments having slowly varying amplitude and predominantly lying along the *b*-axis. They are stacked in the antiparallel manner along the *a*-axis (*Figure 8*). In the $k_1'$ structure, there is also a small out-of-plane component along the *c*-axis.

*Models of the magnetic structures for TbSbTe*

The $k_1$ and $k_2$ propagation vectors in TbSbTe belong to the same special points in BZ as in the case of HoSbTe, X and W respectively. The possible magnetic models were constructed using the ISODISTORT tool. Both vectors have several arms (non-equivalent reciprocal propagation vectors): two (½ 0 0) and (0 ½ 0) for $k_1$, and four (½ 0 ±¼) and (0 ½ ±¼) for $k_2$. It has appeared that for both vectors, the MSGs generated by a single irrep that involve only a single arm of the each of the vectors cannot correctly reproduce the magnetic intensities. Therefore, we have considered the models based on the full star. Here we have an interesting rare case when the multi-arm (multi-*k*) model can be guessed from the diffraction data by using symmetry arguments in the absence of external magnetic field or another external stimulus. The reason for this is that the symmetry of one of the maximal symmetry MSG generated by a primary order parameter irrep allows the presence of secondary irreps that do not break the symmetry of this MSG. Let us first consider the X-point, which allows two irreps $mX_1$ and $mX_2$. The group-subgroup symmetry trees[35, 36] are shown in SI in **Figure S1** and **Figure S2**. In one-arm case the irrep $mX_1$ generates only three magnetic space subgroups $P_amm2$, $P_a2_1/m$ and $P_am$ (basis = {(0,2,0),(1,0,0),(0,0,1)}), of rather high symmetry. None of them is able to fit the experimental intensities. Considering that both arms are involved and again assuming a single primary irrep $mX_1$ we have 10 possible MSG subgroups with three maximal symmetry subgroups $P_c\bar{4}m2$, $P_c4mm$ and $C_a2/m$. On the one hand the full star allows the high symmetry tetragonal MSGs, which cannot be possible in one-arm case. On the the other hand, the symmetry of MSG $C_a2/m$ (basis = {(2,-2,0),(2,2,0),(0,0,1)}) allows the presence of the second irrep $mX_2$, increasing the number of degrees of freedom. The $C_a2$ group (No. 5.17), which is the direct subgroup of the above MSG $C_a2/m$, as well as its subgroups correctly reproduce intensities, having two symmetrically independent Tb sites in the general position (all the components of magnetic moment are free). The overall amplitude on both Tb atoms can be constrained to the same value, yielding a constant magnetic moment structure.

A similar consideration leads to the solution for $k_2$. The one-arm solutions do not have enough free parameters to match the experimental intensities. But already $C_a2/m$ group (No. 12.64, four independent Tb sites), resulting from two-arm star, generated by single irrep $mW_1W_3$ reproduces the magnetic intensities in the unconstrained fit. The attempts to constrain the magnetic moments to the same value substantially worsen the refinement. The subgroup of $C_a2/m$, $C_a2$ (No. 5.17,

basis={(2,-2,0),(2,2,0),(-1,1,2)}) brings four symmetrically independent Tb atoms to a general position with three variable components of magnetic moment on each, which releases enough degree of freedom to constrain the overall amplitude of magnetic moment, again leading to the constant moment structure. This solution can be attained in both $mW_1W_3$ and $mW_2W_4$ physically irreducible representations.

At the base temperature both $k_1$ and $k_2$ phases are present. We refined both models against one diffraction pattern to verify their consistency and the overall picture is provided in *Figure 9*, and the models are plotted in *Figure 10* and *Figure 11*.

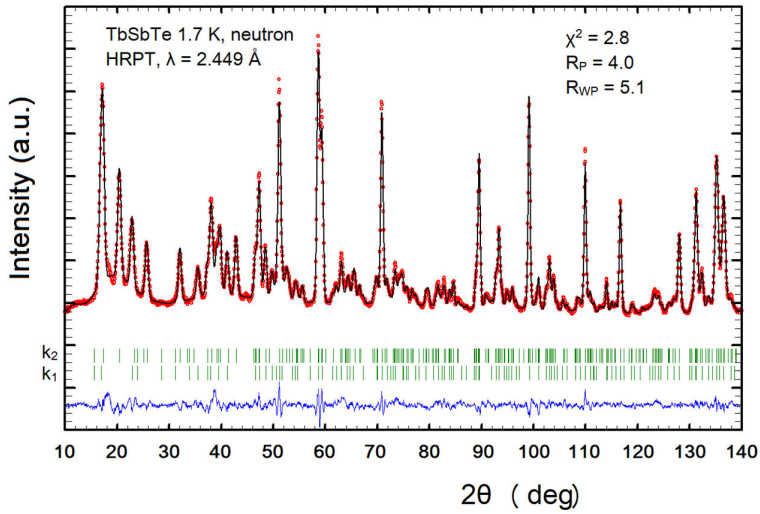

*Figure 9.* Experimental points (red), calculated curve (black), difference curve (blue) and positions of reflections (green) – Rietveld refinement plot for neutron data of TbSbTe at 1.7 K.

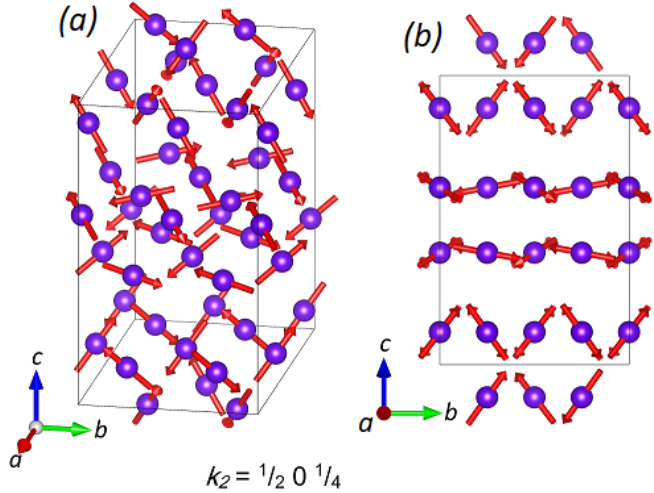

***Figure 10.*** Model of magnetic structure for TbSbTe at 1.7 K, $k_2$ vector: (*a*) a 3D-view, (*b*) projection along the *a*-axis. Only Tb atoms with their magnetic moments are shown for clarity. For interactive models, refer to SI.

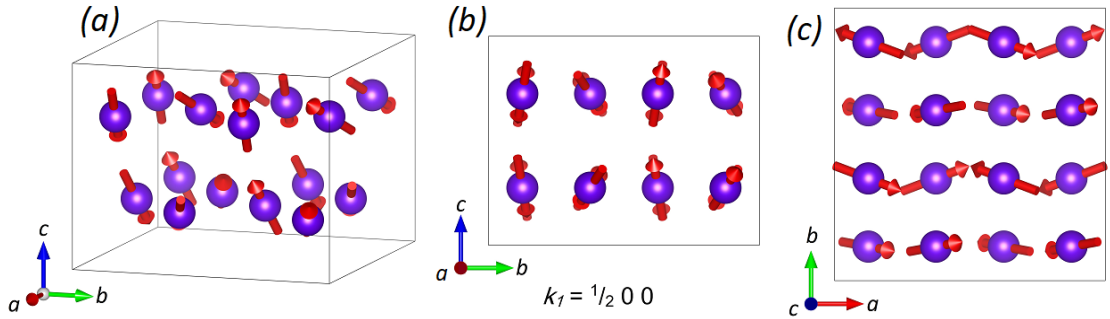

***Figure 11.*** Model of magnetic structure for TbSbTe at 1.7 K, $k_1$ vector: (*a*) a 3D-view, (*b*) projection along the *a*-axis and (*c*) projection along the *c*-axis. Only Tb atoms with their magnetic moments are shown for clarity. For interactive models refer to SI.

In the region, where $k_1'$ and $k_3$ are present, the situation is similar to HoSbTe. Here for $k_1'$ the solution can be found both in mDT$_2$ and mDT$_4$ (*Pmn*2$_1$.1'(00g)*s*0*ss*, No. 31.1.9.2.m124.2). For $k_3$, the best candidate is mU$_2$ (*Pmnm*.1'(0½g)*s*00*s*, No. 59.1.10.7.m406.2). Using the same consideration as for HoSbTe, the magnetic models for $k_1'$ and $k_3$ were refined in separate runs and the summary is provided in ***Figure 12***.

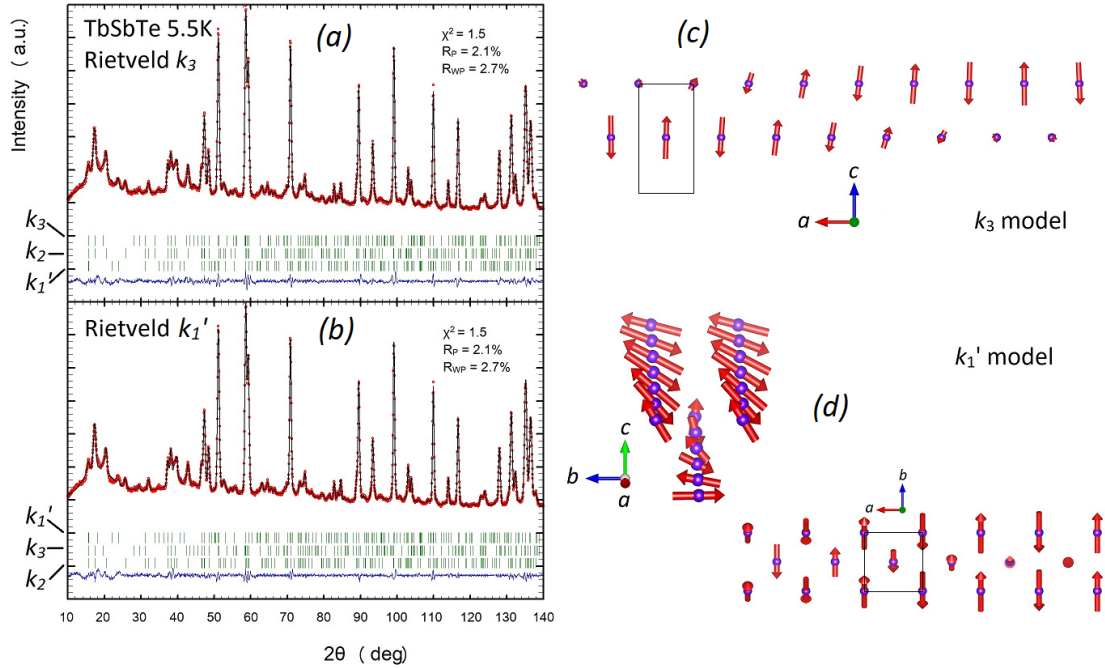

***Figure 12.*** (left) neutron diffraction pattern at 5.5 K for TbSbTe refined using Rietveld method against the $k_3 = (½ - δ\ 0\ ½)$, $δ = 0.035$ model *(a)* and $k_1' = (½ - δ\ 0\ 0)$ model *(b)* (experimental points – red, calculated curve – black, difference curve – blue, positions of reflections – green). The remaining vectors were included in refinement through Le Bail method. (right) the schematic representation of magnetic structures corresponding to $k_3$ *(c)* and $k_1'$ *(d)*.

Unlike for HoSbTe, the magnetic structures for $k_1$ and $k_2$ in TbSbTe are not collinear or coplanar. Each Tb atoms has components of magnetic moment along all three directions as shown in ***Figure 10*** and ***Figure 11***. The magnetic structure build on the $k_3$ vector is essentially the same as for HoSbTe, while the $k_1'$ structure features magnetic moments of variable amplitude revolving in the *a-c* plane.

*General overview and outlook*

Severe overlap of magnetic reflections in powder diffraction patterns is the bottleneck that limits the information that can be extracted from powder data. It is to a lesser extent problematic for the temperature regions where only the commensurate $k_1$ and $k_2$ phases exist, because there is always a significant number of strong well-resolved reflections. Contrary to that, in the incommensurate regime, the overlap might potentially cause ambiguity in interpretation. Still, the fact that we were able to construct physically reasonable high symmetry magnetic models gives credibility to this choice of propagation vectors. The parameters of magnetic structures are provided in ***Table 2***. Neutron *single-crystal* diffraction measurements at different temperatures and fields should allow disentangling the ordering vectors in three dimensions and determination of the relation between the propagation vectors coexisting at each temperature and field point.

The way how the propagation vectors ($k_1$, $k_2$, $k_1'$ and $k_3$) are combined with each other is not quite clear. One can propose two alternative scenarios: *(i)* $k_1$ and $k_2$ form a single phase magnetic structure based on both vectors, where the magnetic moment on a single Ho or Tb is modulated by both vectors simultaneously; *(ii)* $k_1$ and $k_2$ belong to two different magnetic spatially separated phases and these vectors are applied to different atoms.

The first case implies that the scale factor S (or the number of atoms contributing to the specific phase) for all magnetic reflections is equal to the scale factor of the nuclear reflections $S_{k1} = S_{k2} = S_n$, while the second one – that the nuclear scale is shared between two components as $S_{k1} + S_{k2} = S_n$, where $S_{k1} = S_n \cdot \chi_{k1}$, and $\chi_{k1}$ is a volume fraction of the $k_1$-phase.

The first guess on which scenario is realized can be done using the temperature dependence of the magnetic intensities for different vectors. One can notice that for a given vector ($k_1$ or $k_2$), the intensities of magnetic reflections, for example in HoSbTe, raise (for $k_2$) or fade (for $k_1$) with the temperature increase in a coherent manner. This serves as an indication, that within the temperature region where a given propagation vector exists, the underlying magnetic structure remains intact, *i.e.* there is no, for instance, rotation of magnetic moments. This in turn favors the second (multi-phase) scenario, where the temperature effect is expressed just in variations of the volume fraction of each phase. The amplitude of magnetic moment and the scale factor are fully correlated ($S_i \cdot m_i^2$ = constant). Hence, one has to attract additional constrains. Very naturally, it is possible to assume that at the given temperature, the amplitude of magnetic moment in $k_1$ and $k_2$ structures is the same. Applying such a constraint, the value of magnetic moment in HoSbTe at 2.5 K can be refined to 4.985(8)$\mu_B$/Ho with the content of $k_2$ domain equal to 12.8(3) %. For TbSbTe, the amplitude of magnetic moment is 5.84(9) $\mu_B$/Tb and volume fraction of $k_1$ domain is 28.1(3) % at 1.7 K.

*Table 2*. Details of the magnetic structures in HoSbTe and TbSbTe, determined from neutron powder diffraction. Only magnetic atoms are provided; the whole set of structure parameters can be found in SI, *Table 1* and *Table 2*. The profile residuals ($\chi^2$, $R_P$, $R_{WP}$) are shown on the corresponding Rietveld refinement plots. Note that atomic coordinates were fixed to paramagnetic values during the refinement runs. The components of magnetic moment are given in $\mu_B$.

| Atom | x/a, y/b, z/c | $M_x, M_y, M_z$ | \|M\| |
|---|---|---|---|
| | Commensurate $k_1$ and $k_2$ vectors[1] | | |
| | HoSbTe $k_1$ = (½ 0 0), SG $P_a2_1/m$ | | |
| | a = 8.4376(1) Å, b = 4.2188(1) Å, c = 9.1305(1) Å, β = 90° | | |
| Ho | 0.875, 0.25, 0.2754 | 0, 4.985(8), 0 | |
| | HoSbTe $k_2$ = (½ 0 ¼), SG $P_a2_1/m$ | | |
| | a = 8.4376(1) Å, b = 4.2188(1) Å, c = 18.7343(1) Å, β = 103.01(1)° | | |
| Ho1 | 0.1811, 0.25, 0.1122 | 0, 4.985(8), 0 | 4.985(8) |
| Ho2 | 0.4311, 0.25, 0.6122 | 0, 4.985(8), 0 | 4.985(8) |
| | TbSbTe $k_1$ = (½ 0 0), SG $C_a2$ | | |
| | a = 11.9845(2) Å, b = 11.9845(2) Å, c = 9.1986(5) Å, β = 90° | | |
| Tb1 | 0.125, 0.375, 0.7247 | 4.43(6), -2.3(1), 3.03(9) | 5.8(1) |
| Tb2 | 0.375, 0.125, 0.7247 | 3.91(7), 0.76(9), 4.26(7) | 5.8(1) |
| | TbSbTe $k_2$ = (½ 0 ¼), SG $C_a2$ | | |
| | a = 11.9845(2) Å, b = 11.9845(2) Å, c = 19.3291(4) Å, β = 108.06(2)° | | |
| Tb1 | 0.5686, 0, 0.3872 | -4.10(4), 3.11(6), 1.74(6) | 5.8(1) |
| Tb2 | 0.5686, 0.25, 0.8872 | 3.3(2), 1.5(2), 5.70(6) | 5.8(1) |
| Tb3 | 0.8186, 0, 0.8872 | -1.9(1), -3.6(1), 3.70(8) | 5.8(1) |
| Tb4 | 0.8186, 0.25, 0.3872 | 2.1(2), 5.04(6), -1.6(2) | 5.8(1) |

| Atom | x/a, y/b, z/c | $M_{x\text{-}sin}, M_{y\text{-}sin}, M_{z\text{-}sin}$ / $M_{x\text{-}cos}, M_{y\text{-}cos}, M_{z\text{-}cos}$ | \|M\|$_{sin}$ / \|M\|$_{cos}$ |
|---|---|---|---|
| | Incommensurate $k_1'$ and $k_3$ vectors[2] | | |
| | HoSbTe $k_1'$ = 0 0 0.478, SG $Pmn2_1.1'(00g)s0ss$ | | |
| | a = 4.2174(1) Å, b = 9.1273(1) Å, c = 4.2174(1) Å | | |
| Ho | 0, 0.2755, 0.5 | 0, 1.4(3), 2.0(3) | 2.5(4) |
| | | 0, -1.3(3), 2.2(3) | 2.6(4) |
| | HoSbTe $k_3$ = 0 0.5 0.478, SG $Pmnm.1'(0½g)000s$ | | |
| | a = 4.2174(1) Å, b = 9.1273(1) Å, c = 4.2174(1) Å | | |
| Ho | 0.75, 0.2755, 0.25 | 1.295(4), 0, 0 | 1.295(4) |
| | | -1.522(4), 0, 0 | -1.522(4) |
| | TbSbTe $k_1'$ = 0 0 0.465, SG $Pmn2_1.1'(00g)s0ss$ | | |
| | a = 4.2372(1) Å, b = 9.1912(1) Å, c = 4.2372(1) Å | | |
| Tb | 0, 0.2753, 0.5 | 0, 0.4(9), 1.4(2) | 1.4(9) |
| | | 0, 0.8(4), 1.1(6) | 1.1(6) |
| | TbSbTe $k_3$ = 0 ½ 0.465, $Pmnm.1'(0½g)s00s$ | | |
| | a = 4.2372(1) Å, b = 9.1928(1) Å, c = 4.2372(1) Å | | |
| Tb | 0.75, 0.2247, 0.25 | 0 1.363(8) 0.08(3) | 1.37(3) |
| | | 0 -1.599(9) 0.07(2) | 1.60(2) |

[1]The magnetic moment for a given symmetry independent atom at position (x, y, z) is expressed through components ($M_x$, $M_y$, $M_z$) along principle crystallographic axes of the given unit cell. All remaining atoms with their magnetic moments are generated by symmetry elements of the given space group SG.

[2]The magnetic moment at position $\vec{r}(x, y, x)$ is expressed as:
$$\vec{M} = (M_{xsin}\vec{e_x} + M_{ysin}\vec{e_y} + M_{zsin}\vec{e_z}) \cdot \sin(2\pi\vec{k}\vec{r}) + (M_{xcos}\vec{e_x} + M_{ycos}\vec{e_y} + M_{zcos}\vec{e_z}) \cdot \cos(2\pi\vec{k}\vec{r})$$

The derived values of magnetic moments are significantly lower than the expected ordered moments for $Ho^{3+}$ and $Tb^{3+}$ derived, for instance, from Curie-Weiss fit. This is in agreement with the fact, that in the neutron diffraction patterns at the base temperature, the diffuse scattering contribution is rather pronounced (*Figures 3* and *4*), implying that there is a considerable non-long-range ordered component. In the report on macroscopic physical properties of HoSbTe,[20] the authors calculate the magnetic entropy released in the phase transition. It appears to be twice lower, as compared to the values calculated from spin multiplicity. The authors propose that such a deviation might be due to an improper accounting for the lattice contribution in specific heat. Our results offer yet another explanation – that only a fraction of the sample features the long-range order.

Note, that for the incommensurate $k_1'$ and $k_3$ in both compounds we considered only *one-arm solutions*, because they already give very good fit quality. Still, one can make a rather peculiar suggestion for TbSbTe. Since the $k_1$ commensurate magnetic structure is based on two arms of the propagation vector (½ 0 0) and (0 ½ 0) it is natural to suggest that the incommensurate $k_1'$ structure that appears on heating from $k_1$-structure will inherit this *multi-arm* property. Then the MSSG based on the (½ – δ 0 0) and (0 ½ – δ 0) becomes 3D+2. This case is mathematically similar to topological spin structure in CeAlGe[5] (exhibiting the topological Hall effect) with the only significant difference. While in CeAlGe the propagation vector is close to zero ($k$ = (δ 0 0), with δ = 0.065) and the nearest neighbors are aligned nearly ferromagnetically, in TbSbTe, the propagation vector is very close to (½ 0 0), which forces the magnetic moments in neighboring unit cells to be coupled in nearly antiferromagnetic manner with a deviation given by phase shift due to δ. Overall, this hints to the possibility of hosting the topologically non-trivial skyrmion-like phases in our materials. The possible 3D+2 MSSG are provided in SI. The $k_1'$-structure based on both arms of the propagation vector gives very similar fit for our experimental data as the single arm, implying that we cannot distinguish them.

The ordering vectors in metallic systems are predetermined by the nesting points on the Fermi surface and the magnetic exchange parameters, while the actual arrangement of magnetic moments is also dictated by other factors, like crystal anisotropy. In our case, the ground state magnetic structures, being collinear for HoSbTe and non-collinear for TbSbTe, are essentially different, in spite of the fact that they are constructed on the same propagation vectors. One can expect that the electronic structure around the Fermi level is dominated by Sb and Te states, rather than essentially more localized $Ln^{3+}$ states. This together with similar lattice geometry leads to similarities in the Fermi surface across the whole *Ln* series. This explains why both studied members order with the same propagation vectors. The ARPES data for HoSbTe are limited[19], and are not published for TbSbTe. Additional DFT calculations might be useful to relate the features of the Fermi surface and magnetic ordering observed in the title materials.

## *Conclusions*

Our variable temperature neutron powder diffraction studies have revealed several magnetic transitions in magnetically ordered state below seemingly single Néel temperatures in HoSbTe and TbSbTe semimetals, the magnetic members of ZrSiS (PbFCl) structure type (tetragonal space group *P*4/*nmm*). These transitions are identified by the changes of the magnetic propagation vectors and their contribution to the diffraction patterns. The vectors are incommensurate $k_1' = (½ - δ\ 0\ 0)$ and $k_3 = (½ - δ\ 0\ ½)$ slightly below $T_N$ and change to commensurate ones $k_1 = (½\ 0\ 0)$ and $k_2 = (½\ 0\ ¼)$ at lower temperatures. The magnetic structures of the all four phases were determined in maximal possible symmetry magnetic space Shubnikov groups for the commensurate cases and in the magnetic superspace (3D+n) groups for the incommensurate ones. The refined values of magnetic moments within the commensurate phases are: 4.985(8) $μ_B$/Ho and TbSbTe - 5.8(1) $μ_B$/Tb. In the case of TbSbTe we are able to show, that the magnetic structures based on the commensurate $k_1$ and $k_2$ vectors represent a multi-arm case, which may serve as an indication that their counterparts $k_1'$ and $k_3$, being slightly incommensurate, are also multi-arm with a possibility of non-trivial topological magnetic structures. Overall, this work may contribute to understanding the interplay between topological electronic structure and frustrated magnetism in square-net materials.

**Supporting Information:** group-subgroup trees, detailed tables with structure parameters and proposed (3D+2) MSSG (DOC), archive with mcif- and VESTA-files (ZIP).

**Acknowledgements**

The neutron powder diffraction experiments were performed at the Swiss spallation neutron source SINQ, Paul Scherrer Institute (Villigen, Switzerland). We thank the Swiss National Science foundation grants No. 200020-182536/1, 200021_188706 and R'equip Grant No. 461 206021_139082 and SNI Swiss Nanoscience Institute for financial support. We thank Prof. Juan Manuel Perez-Mato and Dr. Vaclav Petricek for valuable discussions.

200021_188706 and R'equip Grant No. 461 206021_139082 and SNI Swiss Nanoscience Institute for financial support. We thank Prof. Juan Manuel Perez-Mato and Dr. Vaclav Petricek for valuable discussions.